\documentclass[conference]{IEEEtran}
\IEEEoverridecommandlockouts
\usepackage{cite}
\usepackage{amsmath,amssymb,amsfonts,mathtools}
\usepackage{algorithmic}
\usepackage{graphicx}
\usepackage{textcomp}
\usepackage{xcolor}
\usepackage{hyperref}
\usepackage{bm}

\usepackage{comment}
\usepackage{listings}
\usepackage{siunitx}
\usepackage{color}
\usepackage{enumitem}
\usepackage{booktabs}
\usepackage{gensymb}
    
\DeclareMathOperator*{\argmax}{arg\,max} %

\newcommand{\PaperTitle}{Mathematical Model of Strong Physically Unclonable Functions Based on Hybrid Boolean Networks}

\begin{document}

\title{\PaperTitle
\thanks{This work was supported by the Department of the Army through award number W31P4Q-20-C-0003. DJG and AP have a financial interest in Verilock.id, which is commercializing the PUF technology described here.}
}

\author{\IEEEauthorblockN{1\textsuperscript{st} Noeloikeau Charlot}
\IEEEauthorblockA{\textit{Physics Department} \\
\textit{Ohio State University}\\
Columbus, OH, USA \\
charlot.5@osu.edu}
\and
\IEEEauthorblockN{2\textsuperscript{nd} Daniel J. Gauthier}
\IEEEauthorblockA{\textit{Physics Department} \\
\textit{Ohio State University}\\
Columbus, OH, USA \\
gauthier.51@osu.edu}
\and
\IEEEauthorblockN{3\textsuperscript{rd} Daniel Canaday}
\IEEEauthorblockA{\textit{Potomac Research, LLC}\\
Alexandria, VA, USA \\
daniel@potomacresear.ch}
\and
\IEEEauthorblockN{4\textsuperscript{th} Andrew Pomerance}
\IEEEauthorblockA{\textit{Potomac Research, LLC}\\
Alexandria, VA, USA \\
andrew@potomacresear.ch}
}

\maketitle

\begin{abstract}
We introduce a mathematical framework for simulating Hybrid Boolean Network (HBN) Physically Unclonable Functions (PUFs, HBN-PUFs). We verify that the model is able to reproduce the experimentally observed PUF statistics for uniqueness $\mu_{inter}$ and reliability $\mu_{intra}$ obtained from experiments of HBN-PUFs on Cyclone V FPGAs. Our results suggest that the HBN-PUF is a true `strong' PUF in the sense that its security properties depend exponentially on both the manufacturing variation and the challenge-response space. Our Python simulation methods are open-source and available at \href{https://github.com/Noeloikeau/networkm}{https://github.com/Noeloikeau/networkm}.
\end{abstract}


\section{Introduction}

\label{intro}
Boolean networks (BNs)\cite{ROSIN,BNS} are connected systems of discrete nodes that exhibit logic-like behavior. They have been used to model gene regulation, design new digital circuits, and have even been studied as a toy model of the human brain. Here, we develop mathematical tools to study a class of BNs that are both challenging to measure experimentally, and useful as cryptography primitives known as Physically Unclonable Functions (PUFs). Specifically we model ultra-fast chaotic circuits known as Hybrid Boolean Networks (HBNs) implemented on Field-Programmable Gate Arrays (FPGAs).

In this context, the nodes of the network are FPGA logic elements, and the edges connecting them are routing resources that conduct the electrical signals representing the analog value of each node state. Unlike standard synchronous designs, nodes exchange signals and perform logical operations continuously - without the control of a global clock or input from the user. Each BN is therefore an autonomous, recurrent electrical circuit that self-oscillates, typically at the maximum frequency allowed by the hardware, which is in the GHz regime for a Cyclone V FPGA. In practice, a clock is used to set the initial state of the system, forming a BN that is a `hybrid' of autonomous and synchronous components (HBN).

These ultra-fast self-oscillations lead to a significant source of entropy generation in HBNs with nonlinear node functions, such as the exclusive-or (XOR) operation. The XOR function is maximally sensitive to individual bit flips, and hence a network of XOR gates can amplify the diverging effects of these bit flips exponentially over time. Thus, chaos has been observed in XOR-HBNs with as little as a single node having two self-loops \cite{BCHAOS2}, and more robustly in a system of three nodes \cite{BCHAOS1}. Chaos is consistently observed above 8 nodes, and 16-node XOR-HBNs have been used as true random number generators with a 12.8 Gbit/s entropy rate \cite{HBNTRNG}.

Recently, XOR-HBNs have been used to develop PUFs on FPGAs \cite{HBNPUF}. PUFs act as `digital fingerprints' by transforming stable environmental information into digital codes used for hardware identification\cite{PUFS}. The atomic-scale manufacturing variations present in the physical properties of FPGA logical elements and wires (such as impedance and length) are amplified and transformed into bits by the dynamics of the HBN. Given an $N$-bit initial condition or \textit{challenge} specifying the binary state of each labeled node in the HBN \textit{class}, the resulting final state or \textit{response} can be used to uniquely identify the individual FPGA or \textit{instance} that created it. A large number of challenge-response pairs (CRPs) is guaranteed, as there are $2^{N}$ possible $N$-bit responses for each of the $2^{N}$ challenges. As a result, the entropy of HBN-PUFs scales super-exponentially with network size, a characteristic of a `strong' PUF.

Due to their super-exponential entropy scaling and sub-nanosecond dynamics, HBN-PUFs stress existing methods of measurement and analysis. Additionally, the FPGA prevents analog readout of the waveform for each node because their primary use is as a digital device. This characteristic significantly limits the amount of information available from experiments. Hence, a model bypassing these limitations is useful for analyzing and certifying the HBN-PUF as an entropy source.

Current HBN models only consider a subset of the realistic experimental conditions present on FPGAs - namely noise on the signal, time-delays on the wires, and differences in logic operation time \cite{LOHMANN,BCHAOS1,BCHAOS2}. Each of these prior works on HBNs have emphasized the importance of variations to these parameters on the resulting dynamics. Here, we develop a mathematical model connecting these parameter variations to the PUF statistics. Specifically, we extend the Glass model \cite{GLASS} to include time-delayed signals \cite{GHIL}, noise, and inhomogeneous time-constants. Our unique contributions follow.

\begin{itemize}
    \item We introduce a mathematical model and Python library for simulating delay-differential network dynamics, specializing to HBN-PUFs on Cyclone V FPGAs. Drawing parameters randomly with means taken from the literature, we find that the model reproduces average experimental HBN-PUF uniqueness and reliability statistics.
    \item The model predicts that PUF uniqueness depends exponentially on the inter-die variation of the FPGA parameters, suggesting that the HBN-PUF is a true `strong' PUF.
\end{itemize}
In Sec. \ref{S1} we introduce the HBN model and define the PUF statistics, then describe our simulation procedure. In Sec. \ref{results} we simulate HBN-PUFs and compare our results to experimental data. We then use the model to estimate the functional form of the PUF statistics in terms of the amount of manufacturing variation and noise. Finally in Sec. \ref{conclusion}, we interpret our results in the context of future work and conclude.

\section{Modeling Framework}
\label{S1}
\begin{figure}[htbp]
\centerline{\includegraphics[width=3.5in]{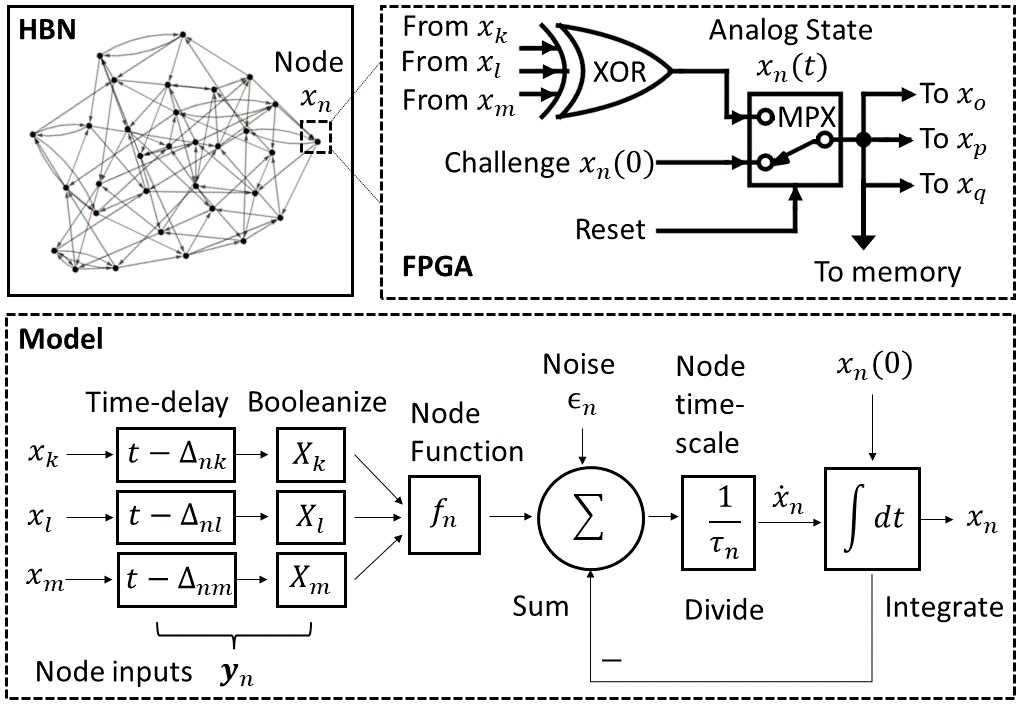}}
\caption{A Hybrid Boolean Network (HBN), depicted as an $N=32$ random regular graph of degree 3, in which each node executes the exclusive-or (XOR) function on the state of 3 random neighbors. The FPGA logic elements and wiring diagram are shown for a single node $n$. The identity of nodes $k-q$ are determined by the fixed topology of the HBN. The block diagram depicts the proposed differential equation \eqref{dxdt} governing the analog node state $x_{n}$.\label{HBNPUF}}
\label{fig}
\end{figure}

We model HBN dynamics as a set of coupled first-order time-delay differential equations with noise, governing the node states $x_{n}\in[0,1]$ for each of the $N$ nodes in the network. This is summarized in Fig. \ref{HBNPUF}. Each node $n$ executes a logic function $f_{n}$ on the states of its neighbors $\bm{y}_{n}$. These neighbor states are time-delayed by $\Delta_{nm}$, representing the finite propagation time required for the state of node $m$ to reach node $n$. These time-delayed signals are also rounded toward 0 or 1, as in hardware logic gates, which convert the incoming analog voltages into binary values. Mathematically,
\begin{equation}
    \tau_{n}\frac{dx_{n}}{dt}=-x_{n}(t)+f_{n}(\bm{y}_{n}(t))+\epsilon_{n}(t),
    \label{dxdt}
\end{equation}
\begin{equation}
    \bm{y}_{n}(t)=\{X_{m}(t-\Delta_{nm})\}_{m\in \text{pred}(n)},
    \label{pred}
\end{equation}
where $X_m(t) = \Theta(x_m(t) - 0.5)$ is the Boolean state and $\Theta$ is the Heaviside step function, and $\text{pred}(n)$ are the predecessors or fixed set of nodes with an edge directed toward node $n$. The characteristic timescale $\tau_{n}$ determines the time required for the node to transition from logic low to logic high or vice-versa, also known as its rise and fall times, respectively. Thermal/charge fluctuations are modeled as Gaussian noise $\epsilon_{n}(t)\sim \mathcal{N}(0,\varepsilon^{2})$ drawn randomly with zero mean and standard deviation $\varepsilon$ at each time-step.

We integrate Eq. \eqref{dxdt} using a first-order Euler method mapping the state from time $t$ to $t+dt$, given by 
\begin{equation}
    x_{n}(t+dt)=x_{n}(t)+\frac{dt}{\tau_{n}}\bigg[-x_{n}(t)+f_{n}(\bm{y}_{n}(t))+\epsilon_{n}(t)\bigg],
    \label{xt}
\end{equation}
where all times and delays are integer multiples of $dt$. The optimal PUF read-out time resides in the transient and is within an order-of-magnitude of the node timescale. Therefore, we integrate for only $\sim10^{1}\bar{\tau}$, and error accumulation due to the first-order approximation is negligible over this interval. Here, $\bar{\tau}=(1/N)\sum_{n}\tau_{n}$ is the mean node time constant. We express all future parameters in units of $\bar{\tau}=0.25$ ns, a value typical of Cyclone V FPGA logic elements \cite{WCD}.

In an actual FPGA, multiplexers are used to set the initial condition of each node in the network as shown in the top right panel of Fig.~\ref{fig}. These multiplexers are held fixed to the challenge for several thousand ns. During this time, the autonomous portion of the network (the XOR gates) stabilizes and synchronously evaluates the challenge $\bm{x}(0)$. Thus, the initial output of each multiplexer $n$ is $f_{n}(\bm{y}_{n}(0))$ (the node function evaluated over its input challenge bits). Experimentally, this is the first value observed in memory.

We observe that simulations explicitly modeling these multiplexers reproduce this behavior. Crucially, we observe that these more complex models produce nearly identical PUF statistics to the simpler models which omit multiplexers but account for this change in initial conditions. Therefore, we omit modeling multiplexers explicitly, and instead use $f_{n}(\bm{y}_{n}(0))$ as the initial condition for each node in the simulation.

\label{model}

HBN-PUF challenge-response pairs (CRPs) are multidimensional arrays having indices summarized in Table \ref{indices}. Classes $s$ are characterized by a particular network topology and set of parameters, representing the circuit netlist instantiated on FPGAs. Instances $i$ of a class are characterized by a uniformly random perturbation of that class's parameters, representing the manufacturing variations present in each FPGA's physical structure. Challenges $c$ are the binary vectors specifying the initial condition of an HBN-PUF instance. Each integration $r$ of an initial condition is characterized by a different sequence of noise vectors randomly generated at each timestep. Each node $n$ then responds with a Boolean state at time $t$. We label this data as $X^{s}_{i,c,r,n}(t)\in\{0,1\}^{N_{s}\times N_{i} \times N_{c} \times N_{r} \times N \times T}$. 

\begin{table}[ht]
\centering
\caption{HBN-PUF challenge-response pair (CRP) indices.}
\begin{tabular}{|c c c|}
\hline
 Index & Symbol & Description\\
 \hline\hline
  Class & $s\leq N_{s}=15$ & Circuit netlist/mean values of $\bm{\tau},\bm{\Delta}$\\ 
 \hline
  Instance & $i\leq N_{i}=8$ & Individual FPGA/exact values of $\bm{\tau},\bm{\Delta}$.\\
 \hline
  Challenge & $c\leq N_{c}=1000$ & Initial condition $\bm{x}(0)$.\\
 \hline
  Repeat & $r\leq N_{r}=100$ & Random noise vector $\bm{\epsilon}(t)$.\\
 \hline
  Node & $n\leq N=256$ & Single bit of $\bm{X}(t)$.\\
 \hline
  Time & $t\leq T=20$ & Register in memory.\\
 \hline
\end{tabular}
\label{indices}
\end{table}

The PUF uniqueness and reliability statistics $\mu_{inter}$ and $\mu_{intra}$ are averaged pairwise differences between responses. Uniqueness measures the average difference between responses of separate instances to the same challenge. It arises due to manufacturing variations, and ideally $\mu_{inter}=0.5$. Reliability measures the average difference between responses of the same instance to the same challenge. It arises due to noise, and $\mu_{intra}=0$ in the ideal noise-free case ($\varepsilon=0$). Let angle brackets denote averages $\langle X_{j} \rangle_{j} = (1/N_{j})\sum X_{j}$, and pairs of indices denote average Hamming distances $\langle X_{j} \rangle_{|j-j'|} = [N_{j}(N_{j}+1)/2]^{-1}\sum_{j< j'\leq N_{j}} |X_{j}-X_{j'}|$. Then,

\begin{equation}
\mu_{inter}^{s}(t) = \langle X^{s}_{i,c,r,n}(t) \rangle_{|i-i'|,c,r,n}
\label{inter}
\end{equation}

\begin{equation}
\mu_{intra}^{s}(t) = \langle X^{s}_{i,c,r,n}(t) \rangle_{i,c,|r-r'|,n}
\label{intra}
\end{equation}

For each class there exists an optimal time $t_{opt}^{s}$ to select the response from the network timeseries. This is when the network has coupled enough to the chip manufacturing variation to be unique, while remaining unaffected enough by noise to be reliable. Mathematically $t_{opt}^s$ is the point maximizing the difference between the statistics
\begin{equation}
    \Delta \mu^{s}(t) : = \mu_{inter}^{s}(t)-\mu_{intra}^{s}(t),
    \label{deltamu}
\end{equation}
\begin{equation}
    t_{opt}^{s}:=\argmax_{t\leq T}\bigg(\Delta \mu^{s}(t) \bigg).
    \label{topt}
\end{equation}

Typically, the class index $s$ is suppressed, as one is dealing with a single PUF class. However, here we will perform an average over classes, representing the typical statistics of a random network ensemble. In these cases an omission of $s$ indicates an average over classes, \textit{i.e.}, $\mu=(1/N_{s})\sum\mu^{s}$.

Simulated HBN-PUF statistics are procedurally generated in the following order: 1. class, 2. instance, 3. CRP. In the first step, class generation, an $N$-node network is drawn from the set of all random regular graphs of degree 3. The delay along each edge is drawn from the Uniform Random distribution over the interval $\Delta^{s}_{nm}\sim\mathcal{U}(0,10\bar{\tau})$. These edge delays represent the location assignment given by the FPGA compiler, which we find can vary from near-zero to several ns depending on the design and routing congestion.

The second step, instance generation, slightly adjusts the parameters of a given class. The node time-constants are $\tau^{s}_{i,n}\sim\mathcal{N}(\bar{\tau},\sigma^{2})$, where $\mathcal{N}$ is the Normal distribution with mean $\bar{\tau}=0.25$ ns and standard deviation $\sigma$. The positive-definite time-delays of instance $i$ are given by $\Delta^{s}_{i,nm}\sim|\mathcal{N}(\Delta^{s}_{nm},\sigma^{2})|$. We interpret $\sigma$ as a uniform manufacturing variation on all FPGA components, and set $\sigma=0.05$ unless otherwise noted, a value typical of modern FPGAs \cite{ManufacturingVariation}. 

The final step, CRP generation, integrates Eqs.~\ref{dxdt}-\ref{xt} to predict the dynamics of the network and formats this data to match the experiment. The integration time $T_{int}=42\bar{\tau}=10.5$ ns and timestep $dt=\bar{\tau}/25=0.01$ ns are fixed for all simulations. The noise amplitude is fixed to the timestep error $\varepsilon=0.01$ unless otherwise noted. For each class, instance, and challenge, \eqref{xt} is integrated $N_{r}$ times using the corresponding parameters. Each integration differs by the unique noise vector drawn as $\epsilon_{n}(t)\sim \mathcal{N}(0,\varepsilon^{2})$. The resulting timeseries is then decimated to match the measurement interval of the experimental data, which is $2\bar{\tau}=0.5$ ns. The first $2\bar{\tau}$ are discarded as they are not observable in experiment, resulting in 20 evenly spaced times over 10 ns from which Eqs.~\ref{inter}-\ref{topt} are calculated.

\section{Results}
\label{results}

\begin{figure}[htbp]
\centerline{\includegraphics[width=3.5in]{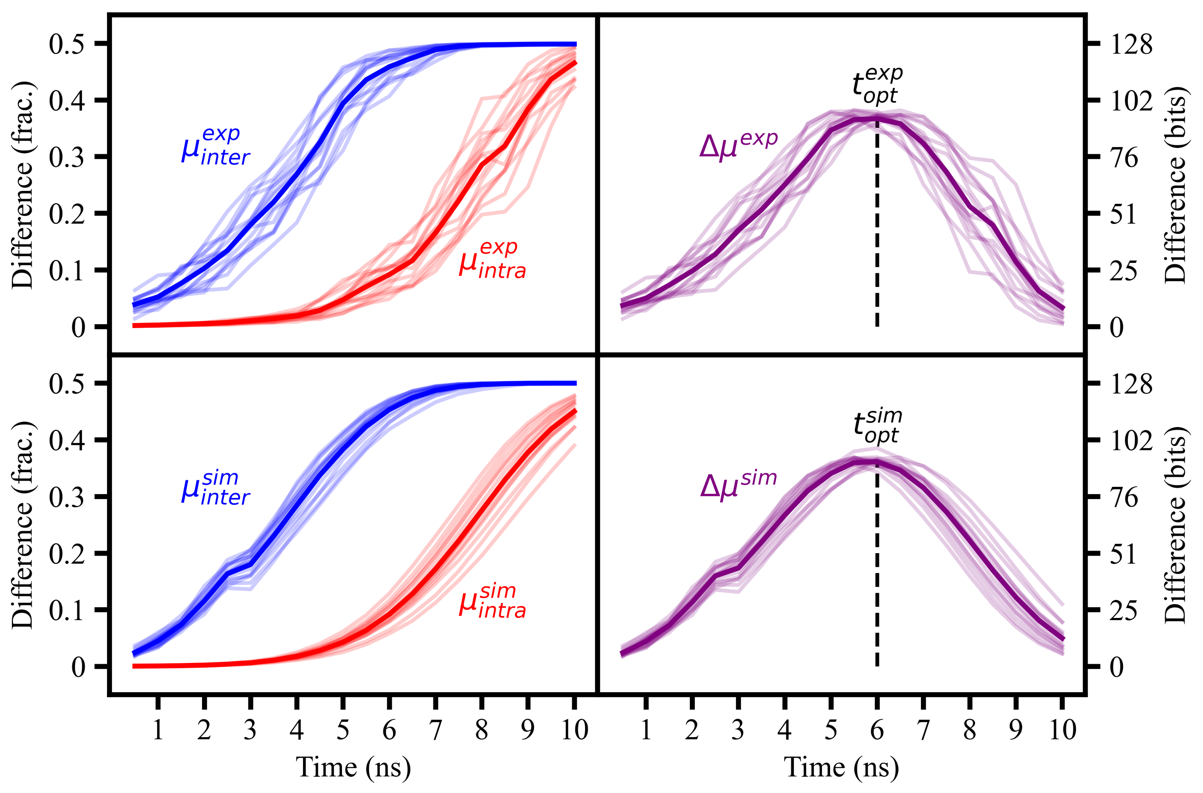}}
\caption{Statistics of experimental (top row) and simulated (bottom row) $N=256$ HBN-PUFs. Light-colored curves are per-class, per-time statistics $\mu^{s}(t)$. Solid curves are an average over $s$, representing a typical HBN-PUF. The sample sizes for the experimental data are given in Table \ref{indices}. The simulated data use a noise amplitude $\varepsilon=0.01$ and manufacturing variation $\sigma=0.05$, with sample sizes $N_{s}=N_{i}=N_{c}=N_{r}=15$. \label{mu}}
\label{fig2}
\end{figure}

Figure \ref{fig2} displays a comparison between actual HBN-PUF experiments on Cyclone V FPGAs (top row) and our model statistics (bottom row). We stress that no attempt is made to fit the parameters of the model to match the experimental data. Nevertheless, the null hypothesis  - that the model does not predict the experiment - should be rejected, due to an RMS Z-score $Z_{RMS}=\sqrt{\langle Z(t)^{2}\rangle_{t}}=0.59<2.33$, where $Z(t)=\frac{ \mu^{sim}(t)-\mu^{exp}(t)}{\sqrt{\text{std(sim)}^{2}+\text{std(exp)}^{2}}} \sim 1 \ \forall t$, and $\text{std(.)}$ are the standard deviations of the $\mu^{s}(t)$. In other words, to within $72\%$ probability (integrated PDF), the ensemble averages of the model and the experiment can be drawn from the same normal distribution. Additionally, the limiting behavior and curvature of both statistics match, demonstrating that the model can be used to make asymptotic predictions about HBN-PUFs.

However, the individual simulated classes and instances display less diversity than their experimental counterparts. This is likely due to differences in the measurement times of nodes arising from skew on the FPGA clock fabric and registers, jitter on the external crystal oscillator, and correlated delays in the routing, none of which we account for in our random parameter draws. All of these factors contribute to the kinks in the experimental curves largely absent in the model. An understanding of these instance-level temporal correlations requires ps-scale intra-FPGA measurement tools such as a waveform capture device \cite{WCD}. Measuring individual $\Delta_{nm},\tau_{n}$ and calibrating the model is the subject of future work.

\begin{figure}[htbp]
\centerline{\includegraphics[width=3.5in]{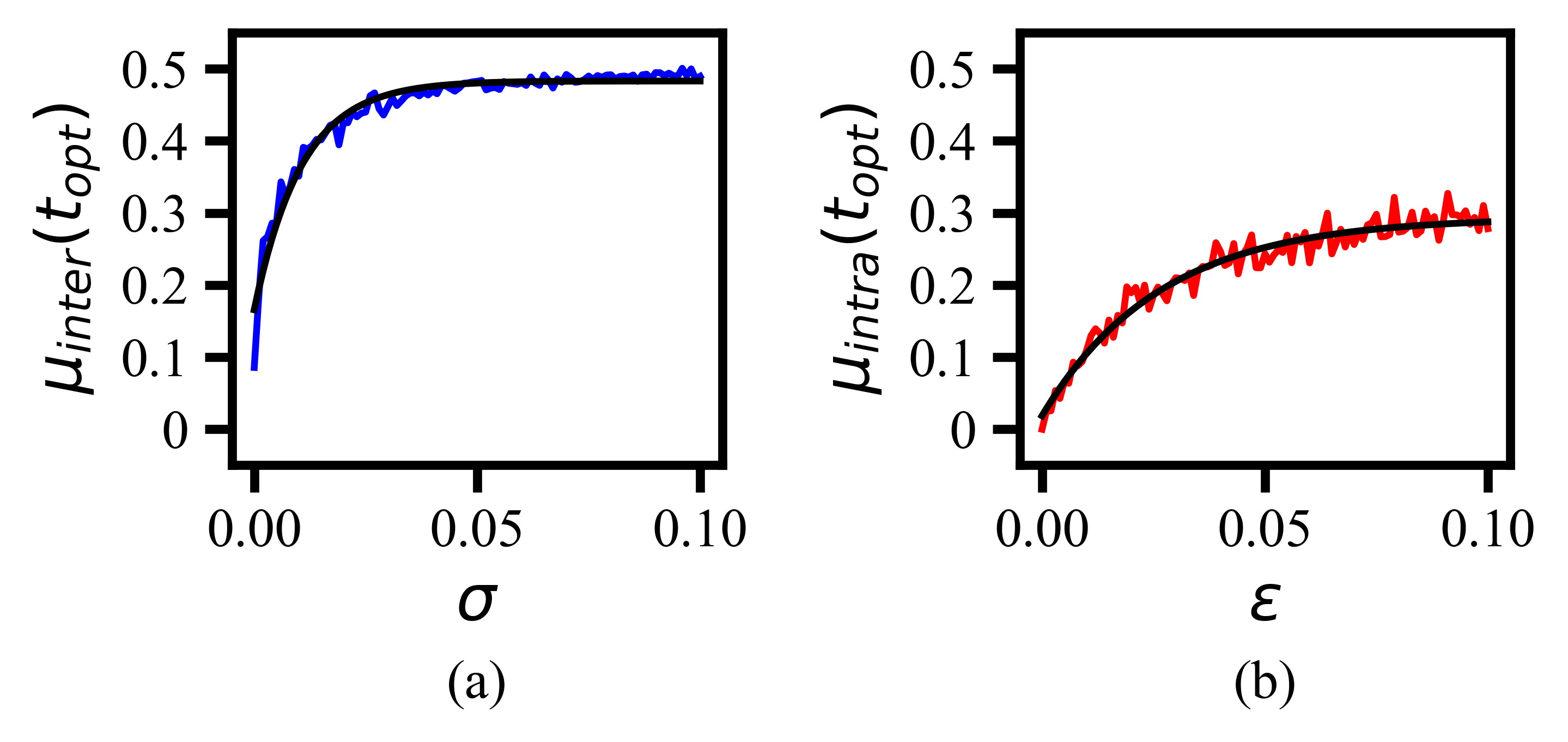}}
\caption{Variation of the average simulated statistics in Fig. \ref{fig2} with respect to $\sigma$ and $\varepsilon$ evaluated at fixed $t_{opt}=6$ ns. The black curves are fits to the function $y=B-Ae^{-Cx}$. (a): Uniqueness as a function of $\sigma$, with a fixed $1\%$ noise floor $\mu_{inter}(\sigma,\varepsilon=0.01)$, having fit parameters $A=0.32\pm0.01$, $B=0.48\pm0.00$, $C=93.78\pm3.98$. (b): Reliability as a function of $\varepsilon$, with a fixed $5\%$ manufacturing variation $\mu_{intra}(\varepsilon,\sigma=0.05)$, having fit parameters $A=0.28\pm0.01$, $B=0.29\pm0.00$, $C=36.67\pm2.27$.}
\label{fig3}
\end{figure}

Next, we determine the extent to which the HBN-PUF model scales with manufacturing variation and noise, and how this information quantifies the PUF strength. To this end, we calculate the statistics $\mu_{inter}$ and $\mu_{intra}$ as a function of $\sigma$ and $\varepsilon$ by fixing the observed optimal response time $t_{opt}=6$ ns. These data are shown in Figure \ref{fig3}. 

In both cases, the data are well fit to a saturating exponential function. This provides mathematical evidence suggesting that the HBN-PUF is truly `strong'. This is because, in addition to having an entropy that scales super-exponentially with $N$, the HBN-PUF uniqueness statistic $\mu_{inter}$ scales exponentially with the inter-die manufacturing standard deviation $\sigma$. Furthermore, the exponential rate constant $C$ fit to $\mu_{inter}(\sigma)$ is roughly three times that of $\mu_{intra}(\varepsilon)$. This demonstrates that the nonlinear dynamics amplify manufacturing variations to a greater degree than noise, ensuring effective HBN-PUF behavior. Additionally, exponential scaling of $\mu_{inter}(\sigma)$ indicates that small uncertainties in the system parameters result in large deviations of the modeled response. Log-$\sigma$ simulations (not shown) demonstrate that this uncertainty conservatively approaches the observed noise floor at $\sigma_{min}=10^{-3}$, \textit{i.e.}  $\mu_{inter}(\sigma_{min}, \epsilon=0.01) \approx \mu_{intra}(\sigma=0.05, \epsilon=0.01)$. We therefore estimate that the dynamics of each node and edge in the HBN-PUF must be resolved to $\sigma_{min}\bar{\tau}\approx 10^{-13}$ s (sub-ps-scale) precision in order for a modeling attack to be feasible. FPGA clock fabrics routinely require ns-scale skew for large recurrent networks like the HBN. However, small, carefully placed networks may be measurable to ps-scale and below over short time periods, and this is the subject of future research.

\section{Conclusions and Future Work}
\label{conclusion}
In summary, we accurately predict the class-averaged experimental statistics of HBN-PUFs using stochastic first-order time-delay differential equations and parameter values drawn randomly from the literature. These simulations predict that HBN-PUFs are exponentially sensitive to manufacturing variations to a greater degree than noise. In conjunction with the exponential entropy scaling, our results suggest that the HBN-PUF is a true `strong' PUF. The introduced simulation library supports arbitrary network topologies and logical functions, and in the future we will combine it with $5$-ps-resolution experimental measurements using a waveform capture device on the Cyclone V\cite{WCD}. These measurements will be used to test the predicted sub-ps uncertainty bound required for a modeling attack. We will also present novel theoretical metrics analyzing strong PUF properties, specifically unclonability, entropy, and multidimensional chaos. 


\end{document}